\documentclass[preprint,authoryear,12pt]{elsarticle}%
\usepackage{graphicx}
\usepackage{epsfig}
\usepackage{amssymb}
\usepackage{amsthm}
\journal{New Astronomy}
\begin{document}
\begin{frontmatter}
\title{X-ray emission from O-type stars : DH Cep and HD 97434}
\author[ari]{Himali Bhatt\corref{cor1}}
\ead{himali@aries.res.in}
\author[ari]{J. C. Pandey\corref{cor2}}
\ead{jeewan@aries.res.in}
\author[ari]{Brijesh Kumar}
\ead{brij@aries.res.in}
\author[ari]{Ram Sagar}
\ead{sagar@aries.res.in}
\address[ari]{Aryabhatta Research Institute of Observational Sciences, Nainital, India 263 129}
\author[tif]{K. P. Singh}
\address[tif]{Tata Institute of Fundamental Research, Mumbai, India 400 005}
\ead{singh@tifr.res.in}

\begin{abstract}
We present X-ray emission characteristics of the massive O-type stars DH Cep
and HD 97434 using archival XMM-Newton observations.
There is no convincing evidence for short term variability in the X-ray intensity during the observations.
However, the analysis of their spectra reveals X-ray structure
being consistent with two-temperature plasma model. The hydrogen column
densities derived from X-ray spectra of DH Cep and HD 97434 are in agreement
with the reddening measurements for their corresponding host clusters NGC 7380 and
Trumpler 18, indicating that the absorption by stellar wind is negligible.
The X-ray emission from these hot stars is interpreted in terms of the
standard instability-driven wind shock model.
\end{abstract}


\end{frontmatter}

\section{Introduction}
\label{sec:intro}

Massive, hot and luminous O-type stars have powerful stellar winds, with mass loss rates of the order of 10$^{-6}$ to 10$^{-4 }$ M$_\odot$ yr$^{-1}$ and velocities of several thousand km s$^{-1}$. These winds play a crucial role both in the stellar evolution as well as the galactic evolution. The existence of strong winds is supposed to be a primary reason for X-ray emission in the early type stars. Study of the X-ray emission from O-type stars is important for understanding the high energy processes that generate hot gas in the wind or in the  atmosphere of hot stars.  X-ray emission from O-type stars  was first discovered serendipitously with the EINSTEIN observatory (Harnden et al. 1979).  Most O-type stars are reasonably bright ($\rm{10^{31} < L_{X} < 10^{34}~erg~s^{-1}}$ ) and have a soft thermal spectrum with kT $<$ 1 keV. The X-ray luminosity of these sources is found to scale with their bolometric luminosity as $\rm{L_X /L_{bol}} \sim 10^{-6}$ to 10$^{-8}$ (Long \& White 1980, Pallavicini et al. 1981, Singh \& Naranan 1982, Bergh$\rm{\ddot{o}}$fer et al. 1997; Sana et al. 2006). It is generally thought to be produced as a result of shocks, with velocity jumps up to a few hundred km s$^{-1}$ , generated throughout the stellar wind because of dynamic instabilities (Lucy \& White 1980; Owocki \& Cohen 1999; Kudritzki \& Puls 2000).  This scenario is now called a standard model for X-ray emission from early-type stars (Rauw 2008).  Although, most of the X-ray emission is dominated by soft X-rays as expected from wind-shock model (e.g. Feldmeier, Puls \& Pauldrach 1997), a significant and an unexpected harder component is also seen in the X-ray spectra (e.g. Leutenegger \& Khan 2003; Bross et al. 2007; Stelzer et al. 2005).  High quality observational data from XMM-Newton and Chandra observatories are indeed crucial to provide a strong test of the latest models. In order to understand the X-ray emission processes from O-type stars, we have analyzed XMM-Newton data for two O-type stars DH Cep and HD 97434. The results of our analysis are presented here.

DH Cep is an O6 V + O7 V spectroscopic binary system and a member of the cluster NGC 7380 (distance = 3.7 kpc, age = 10 Myr; Underhill 1969; Massey, Johnson \& De Gioio-Eastwood 1995; Hilditch, Harries, \& Bell 1996).  It has an orbital period of 2.11 d (Penny, Gies \& Bagnuolo 1997). A low signal, Einstein X-ray observation is reported by Chlebowski, Harden \& Sciortino (1989) with
$\rm{L_X\sim5.37\times10^{33}~erg~s^{-1}}$ in the energy band 0.2--3.5 keV.

HD 97434 is an O9 V star (Skiff 2007), with a membership probability of 78\%
(Dias et al. 2006) in the cluster Trumpler 18 (distance = 1.5 kpc; age = 15 Myr;
V$\rm{\acute{a}}$zquez \& Feinstein 1990). The Einstein X-ray data
for the source are cataloged by Chlebowski, Harden \& Sciortino (1989)
with $\rm{L_X<3.4\times10^{32}~erg~s^{-1}}$ (corrected for the distance of 1.5 kpc) in the energy band 0.2--3.5 keV.

\section{XMM-Newton Observations and Data Reduction}
\label{sec:reduc}
We present analysis of the archival data obtained with the XMM-Newton
observatory.
Table~\ref{tab:xray_observations}  summarizes the XMM-Newton observations.
Our analysis is based on CCD images, lightcurves and spectra from the
European Photon Imaging Cameras (EPIC). Data were acquired simultaneously with
EPIC-PN camera (Str$\rm{\ddot{u}}$der et al. 2001) and two nearly identical EPIC-MOS
(MOS1 and MOS2;Turner et al. 2001) cameras. Data reduction followed standard procedures using XMM-Newton Science Analysis System
software (SAS version 8.0.0) with updated calibration files (Ehle et al. 2004).
 A detailed description of the data reduction and analysis procedure is
given in Bhatt et al. (2010). Owing to poor counts below 0.3 keV and above 7.5 keV, our analysis is limited
to the energy band 0.3--7.5 keV.
 For the star HD 97434, the observations are not made with the MOS1 detector, therefore
we used  only the data from PN and MOS2 detectors for further analysis.
For the star DH Cep, the observations have been taken during the binary phase of 0.08--0.25. The binary phases of DH Cep were determined using the ephemeris HJD = 2441905.805+2.110932E (Hilditch, Harries, \& Bell 1996; Penny, Gies \& Bagnuolo 1997).

Source photons were extracted from circular region of 20$^{\prime\prime}$ around the source position
to generate the light curves and spectra of the source DH Cep and HD 97434.
Background was estimated from a number of empty regions close to the X-ray source
in the detector.
X-ray spectra of the sources were generated using SAS task {\sc especget}, which
also computes the photon redistribution as well as the ancillary matrix.
Finally, the spectra were
re-binned to have at least 20 counts per spectral bin for both the sources.

\section {Analysis}
\label{sec:ana}
\begin{figure*}
\includegraphics[width=3.0in]{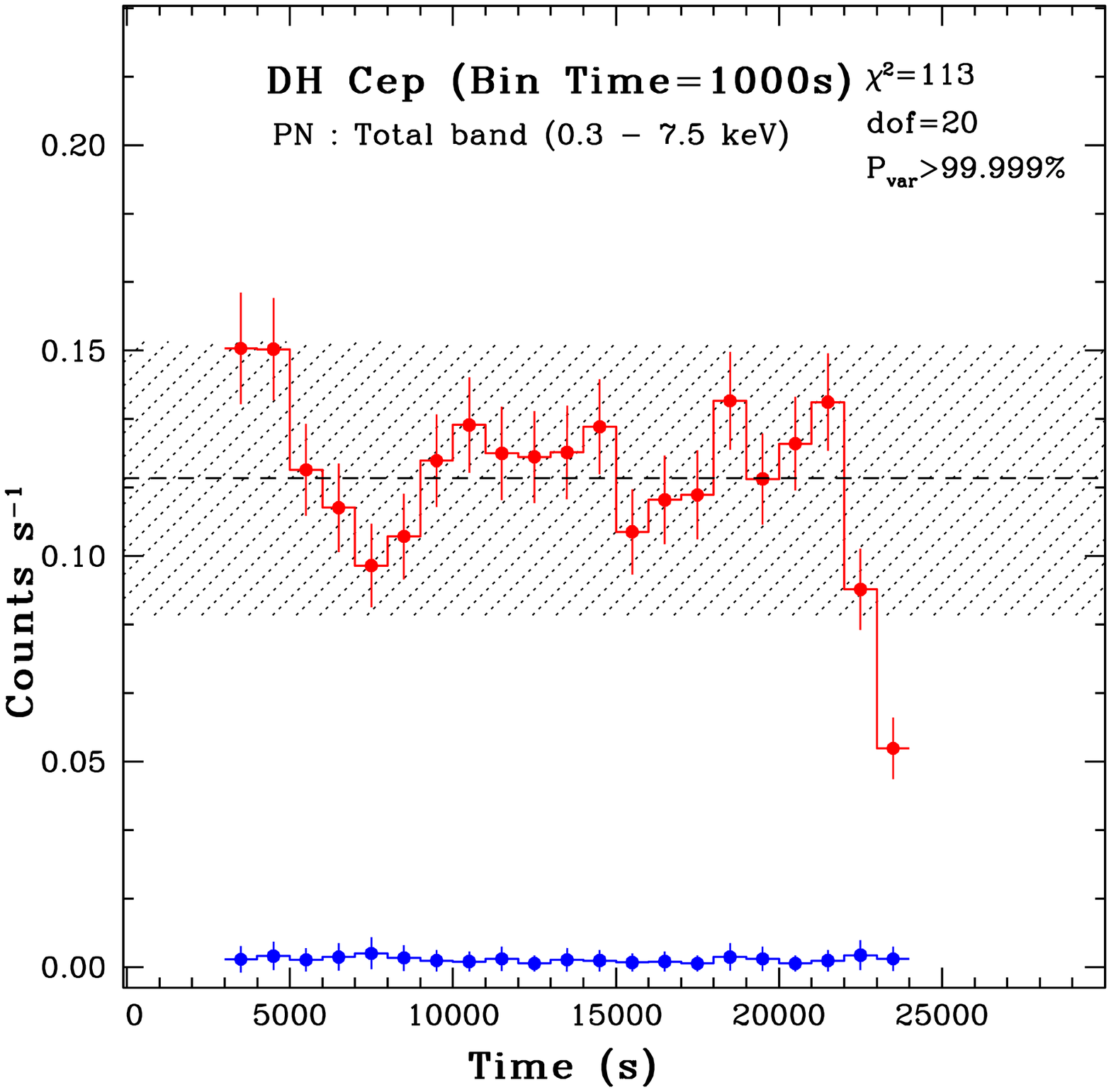}
\includegraphics[width=3.0in]{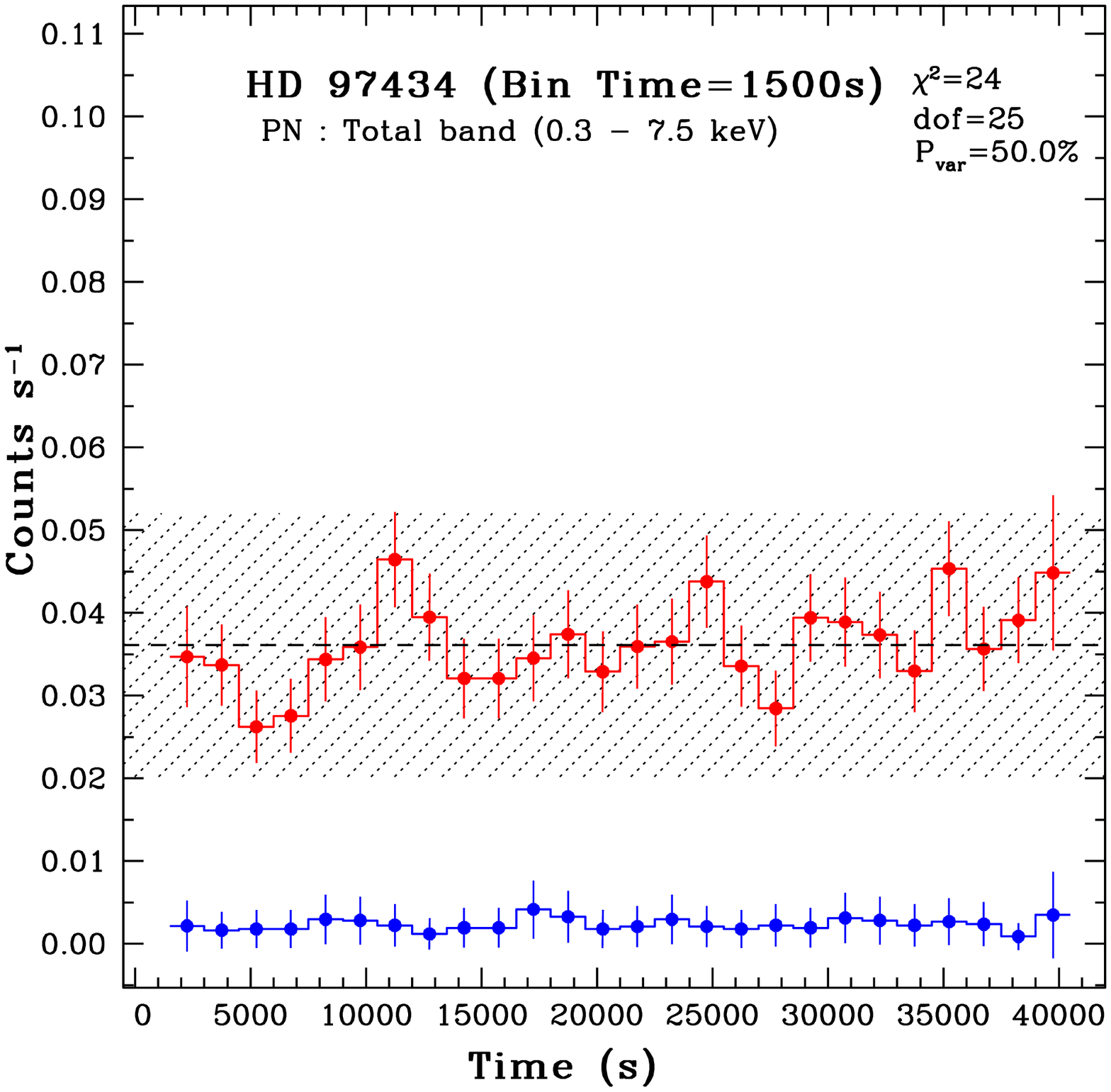}
\includegraphics[width=3.0in]{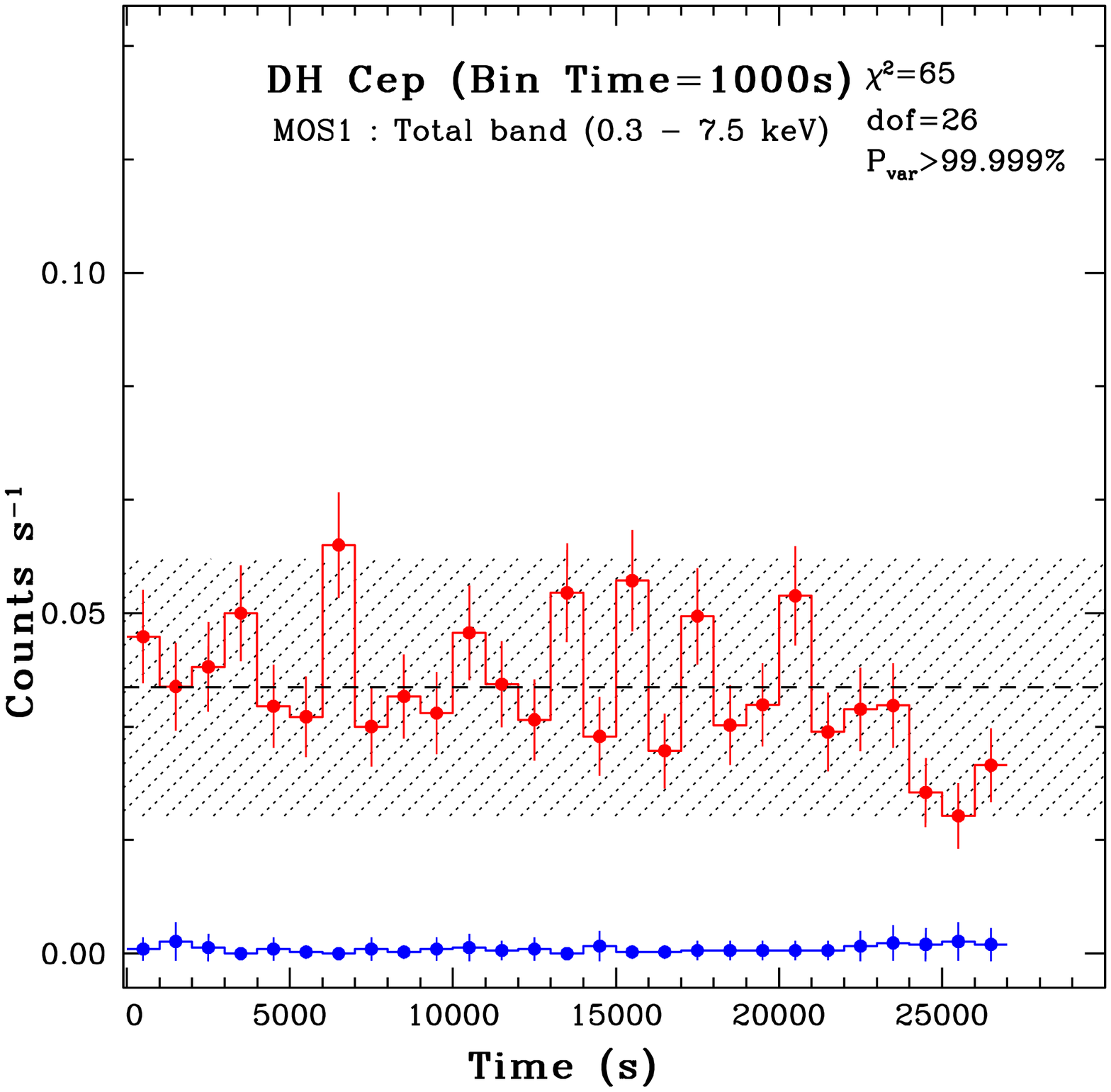}
\includegraphics[width=3.0in]{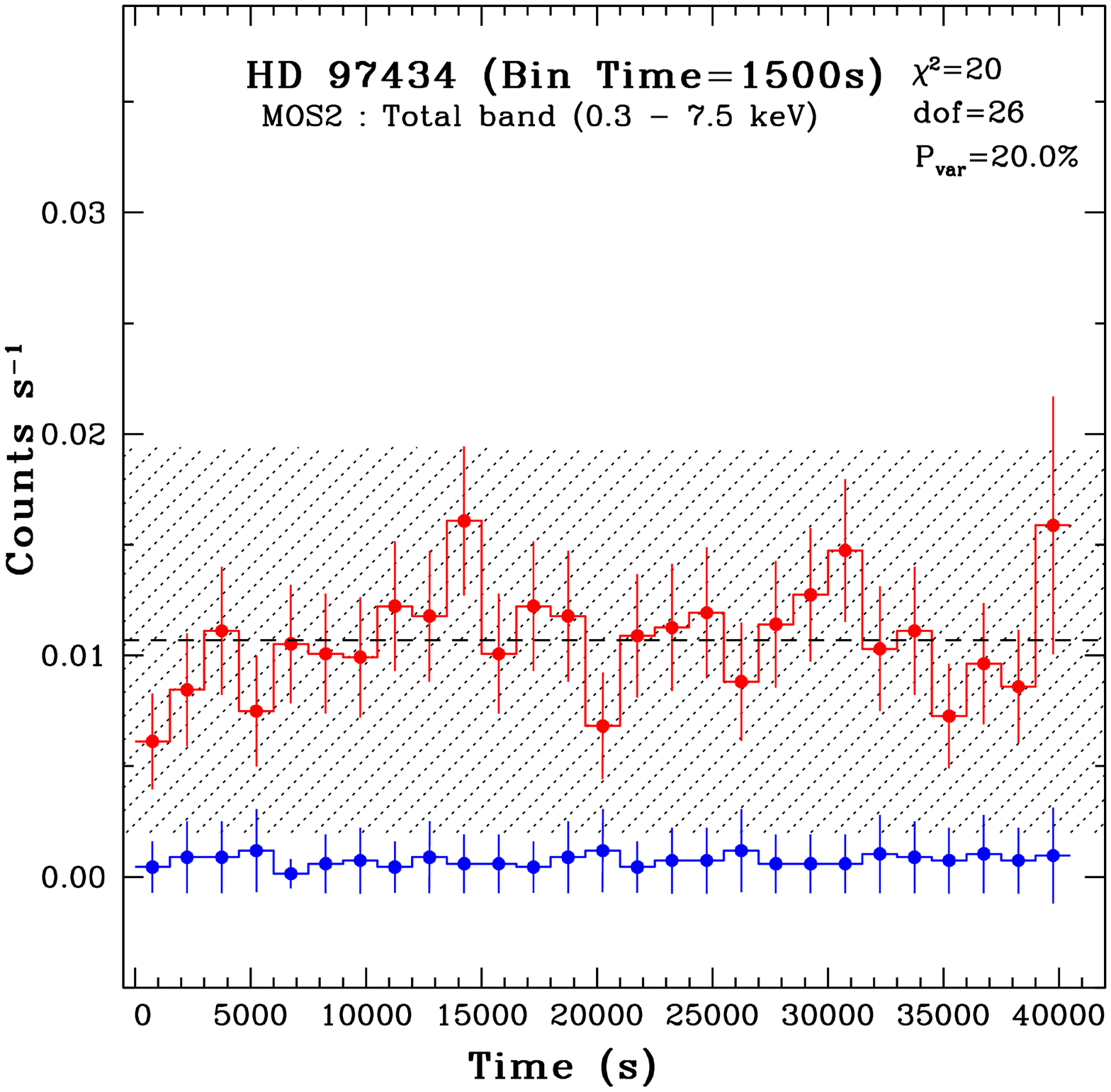}
\includegraphics[width=3.0in]{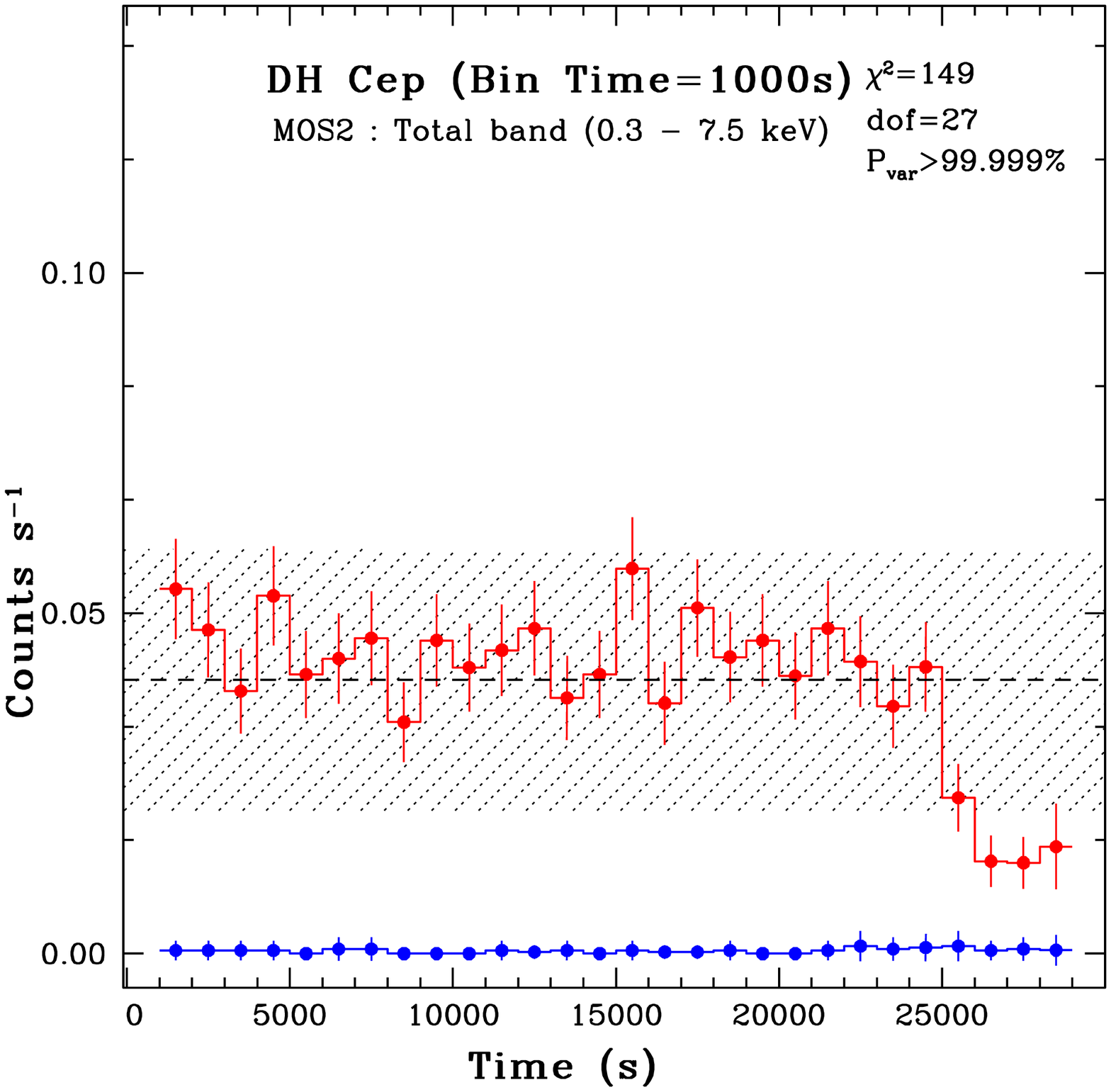}
\caption{X-ray light curves for the stars DH Cep (left panel) and HD 97434 (right panel).
The background subtracted lightcurves are shown by top set of circles and  the background in the same XMM-Newton/EPIC field of view represented
by bottom set of circles in each panel.
}
\label{fig:mass_lt}
\end{figure*}

Figure~\ref{fig:mass_lt} shows the background and the background subtracted X-ray lightcurves of the stars DH Cep and HD 97434. We have binned the event lists into 1000 and 1500s time intervals for the stars  DH Cep and HD 97434, respectively.  We have performed  $\rm{\chi^2}$ test to measure the significance of the deviations from the mean count rates. The probability of the variability ($\rm{P_{var}}$) to
quantify the constancy of the data over the time scale of observations was calculated. The results are displayed in the respective panels of Figure~\ref{fig:mass_lt}.  The value of $\rm{P_{var}}$ was calculated to be $>99$\% for the star DH Cep. However, it was calculated to be $< 50$\% for the star HD 97434. This indicates that the light curve of the star DH Cep shows significant variability.
We further analysed the light curves of DH Cep and found that the observed variability may arise due to a few outlier points. If we neglect the four outlier points of the light curve of DH Cep, the  probability of variability decreases to 80\%.  Therefore, we have plotted the $\rm{3\sigma}$ limit  in Figure~\ref{fig:mass_lt} represented by a shaded region around the constant count rate, i.e., mean value of count rate for the data sample.  Here, $\rm{\sigma}$ represents the standard deviation in count rates.  This analysis shows that none of the data points in any of the light curves were lying outside of $\rm{3\sigma}$ limit.  Further, the variability trend seen in the PN data was not correlated with the MOS data and all the data points in the light curves were within $\rm{3\sigma}$ limits.
Therefore, it appears that the significance of the variability in the light curves is very poor and cannot be detected by the present data set.

\begin{figure*}
\hbox{
\includegraphics[width=2.2in, angle=-90]{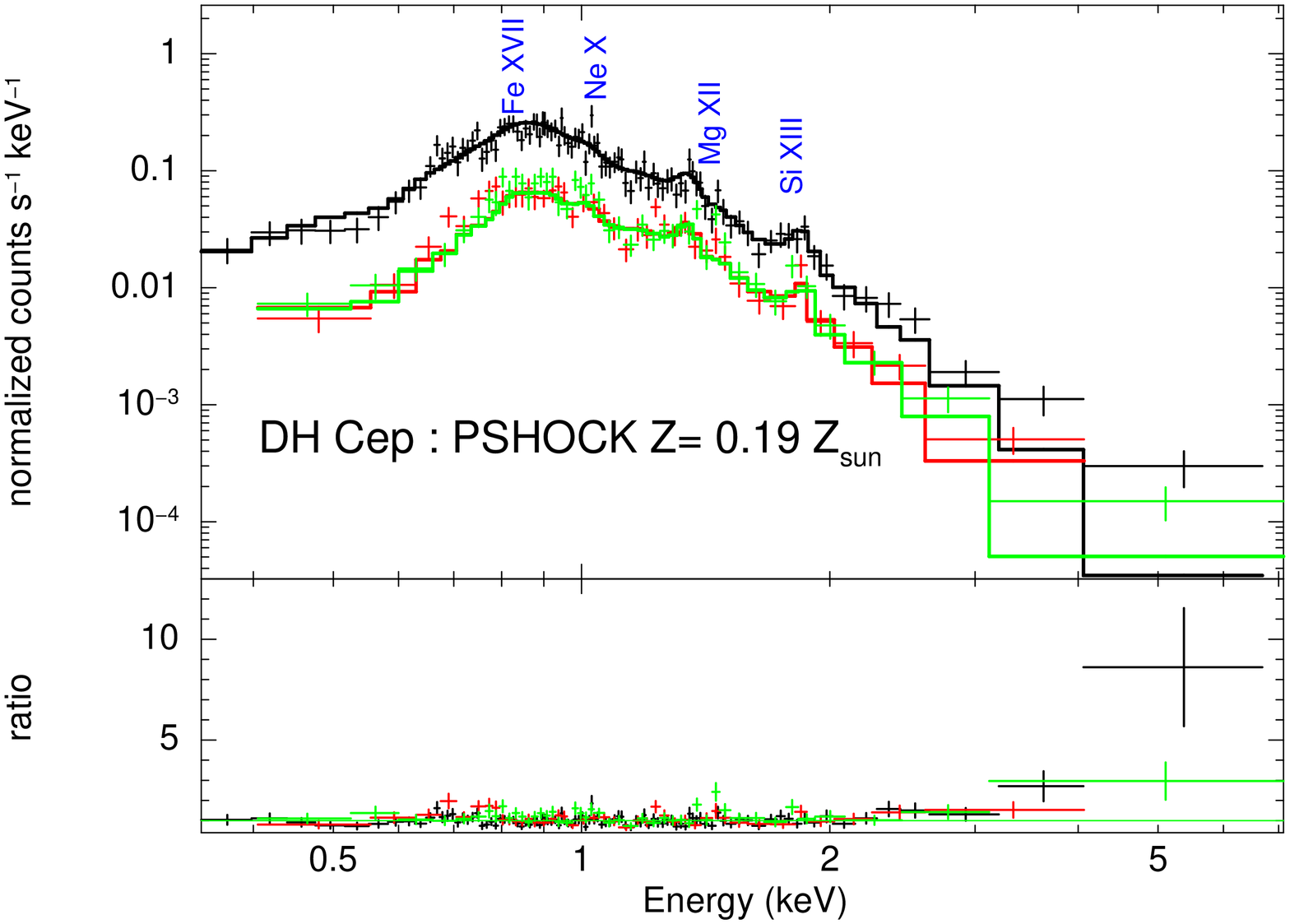}
\includegraphics[width=2.2in, angle=-90]{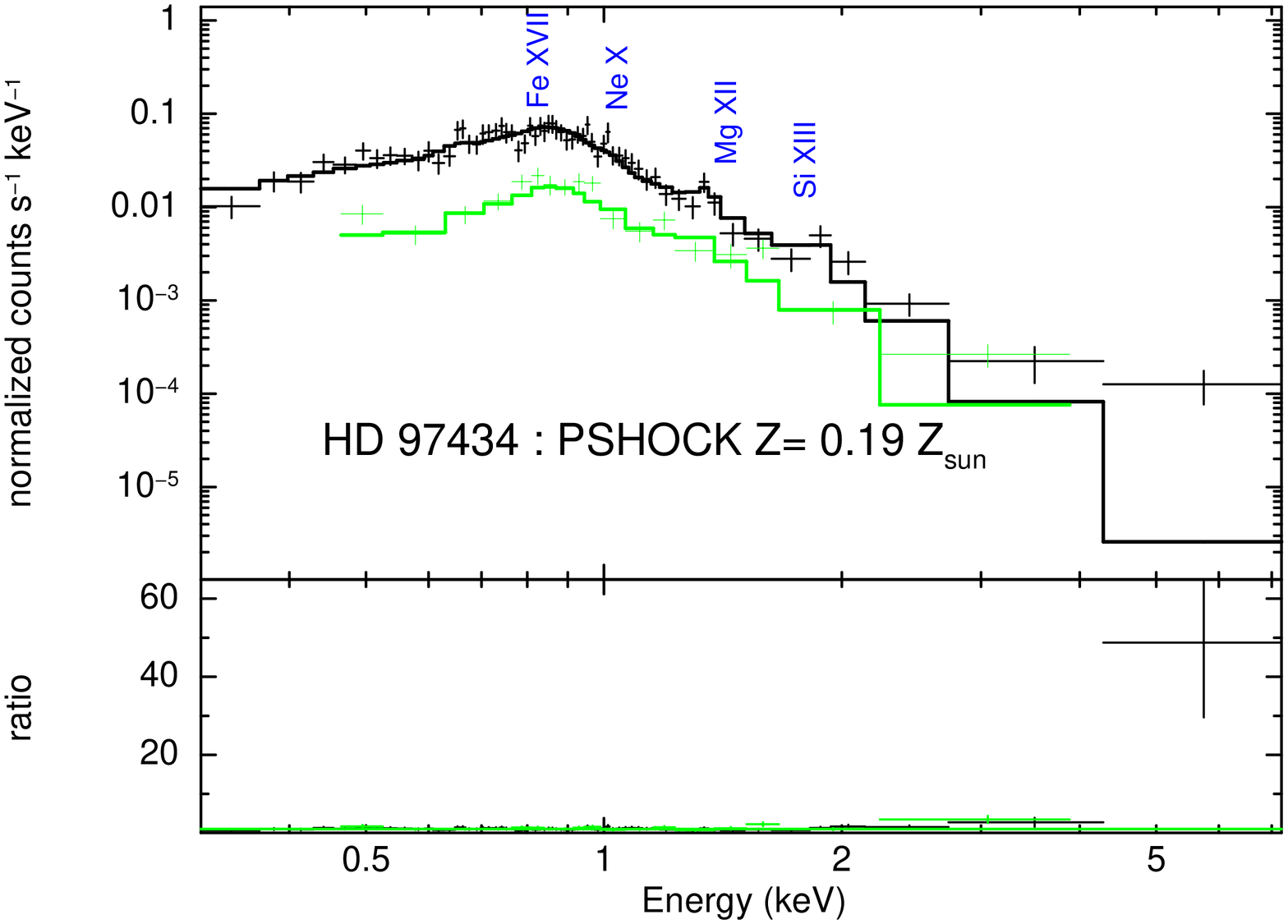}
}
\hbox{
\includegraphics[width=2.2in, angle=-90]{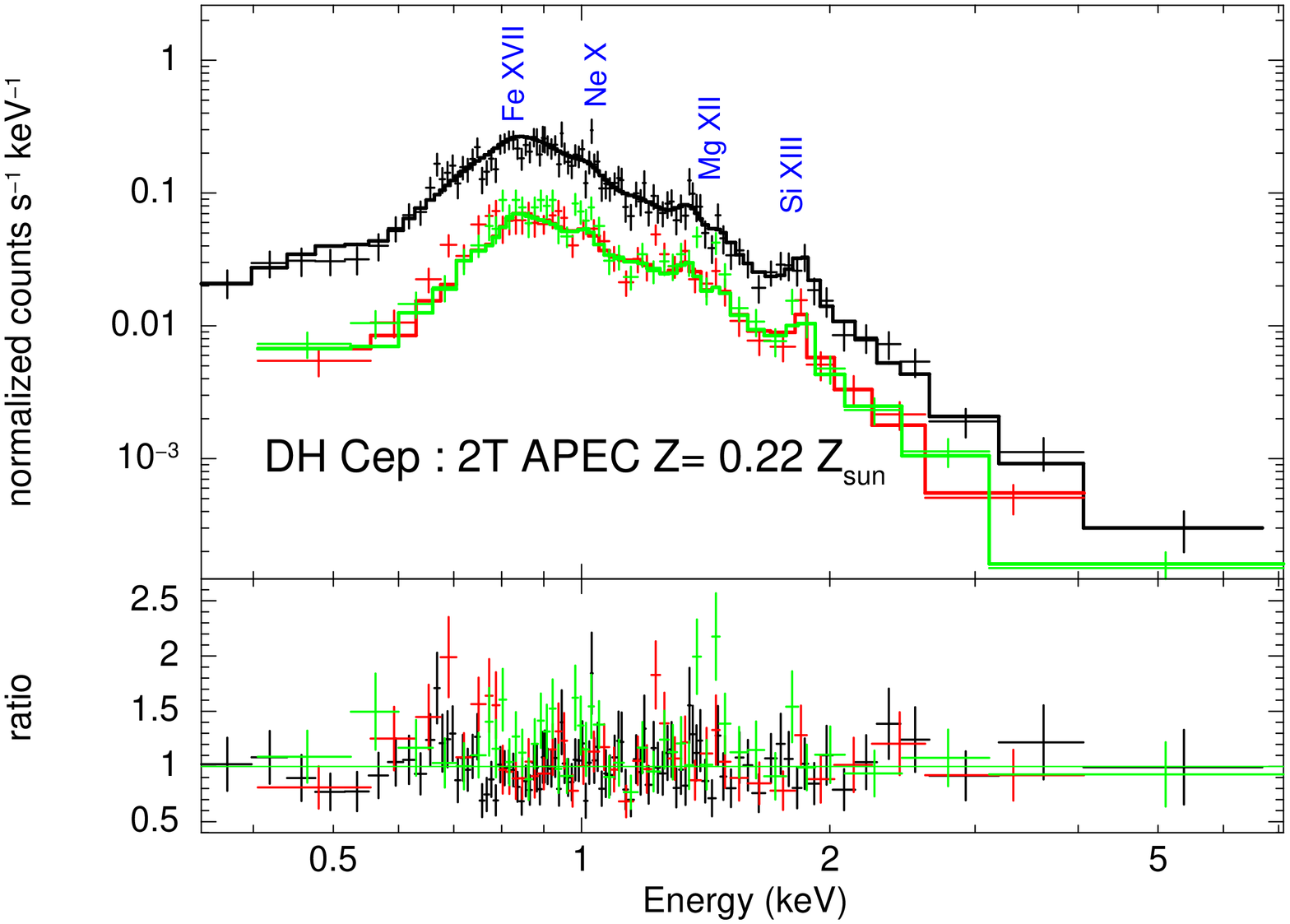}
\includegraphics[width=2.2in, angle=-90]{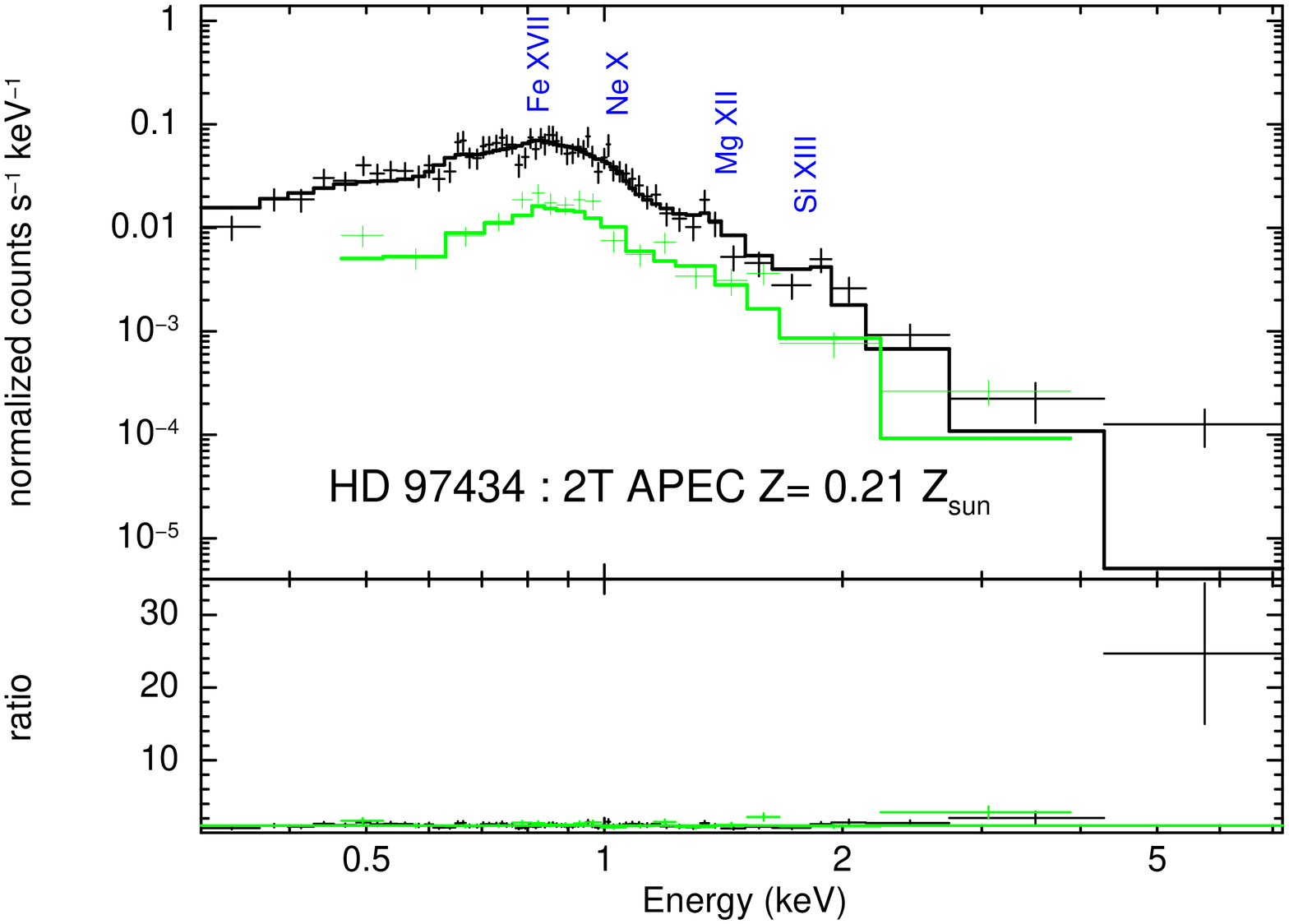}
}
\caption{Best-fit  X-ray spectra of  massive stars. left panel: for DH Cep.
right panel : for HD 97434. The solid line histograms represent the best fit {\sc pshock} (in upper panels) and 2T {\sc apec} model (in lower panels)
with absorption model {\sc PHABS} for PN, MOS1 and MOS2 data.  }
\label{fig:mass_spt}
\end{figure*}

X-ray  spectra of DH Cep and HD 97434 are shown in Figure~\ref{fig:mass_spt}.
Emission lines of Fe XVII (0.8 keV), Ne X (1.02 keV), Mg XII (1.47) and Si XIII (1.853 keV)
are seen in their respective spectra.
X-ray emission from massive stars is thought to be produced from thermal emission,
 generated from shock-heated plasma. These hydrodynamic shocks may occur
in the stellar winds from massive stars. However, in massive binaries they may occur in
wind-wind collision zone (Bhatt et al. 2010 and references therein).
Therefore, we have modeled the spectra using
 (a) plane-parallel shock model ({\sc pshock}; Borkowski, Lyerly \& Reynolds 2001),
and (b) models of Astrophysical Plasma Emission Code
({\sc apec}; Smith et al. 2001), as implemented in the XSPEC version 12.3.0.
A $\chi^2$ -- minimization  gave the best fitted model to the data.
We corrected for the local  absorption in the line-of-sight to the source ($\rm{N_H}$) using
the photoelectric absorption cross
sections according to Baluci$\rm{\acute{n}}$ska-Church \& McCammon (1992) and
modeled it as {\sc phabs} (photoelectric absorption screens).
The abundances in the absorber were allowed to vary during the fitting procedure
The abundances were varied with respect to solar abundances adopted from Lodders (2003).

First, we fitted {\sc pshock} plasma model to derive their spectral features. The constant temperature {\sc pshock} plasma model was considered without incorporating mass-loss and orbital parameters of massive stars.  However, the model does account for non-equilibrium ionization effects and assumes an equal electron and ion temperature. The best fit {\sc pshock} models to the data  are shown in left and right top panels of Figure \ref{fig:mass_spt} and the best fit parameters are given in Table~\ref{tab:spec_massive_bestfit}.  The best-fit X-ray temperatures deduced from X-ray spectra are found to be $\rm{0.64^{+0.02}_{-0.02}}$ keV and $\rm{0.64^{+0.04}_{-0.04}}$ keV for DH Cep and HD 97434, respectively.

Second, the simplest spectral {\sc apec} model
was used  for the spectral fitting. Here, we first considered the case of an isothermal
hot ionized gas as the one temperature (1T) {\sc apec} model.
This model is expressed as {\sc phabs $\times$ apec}.
The 1T {\sc apec} models with either solar or subsolar photospheric abundances were found to be unacceptable for both the stars
due to high value of $\rm {\chi_{\nu}^{2}}$.
Therefore, for the next level of complexity, we used
models with two temperatures (2T) expressed as
{\sc phabs(apec+apec)}.  In 2T models the first component
represents the "cooler" ionized gas and the second component  represents the "hotter" ionized gas.
The 2T models were found to be acceptable but required subsolar abundances, i.e., $\rm{Z\approx0.2~Z_\odot}$.
The subsolar abundances have also been reported  for 15 other massive stars by Zhekov \& Palla (2007).
The best fit 2T models are displayed
as histograms along with the data in the left and the right bottom panels of Figure~\ref{fig:mass_spt}
for the stars DH Cep and HD 97434, respectively.
The best-fit parameters for these models were obtained by $\chi^{2}$-minimization technique
and are given in Table~\ref{tab:spec_massive_bestfit} with their corresponding $\chi^{2}_\nu$.
We noted that 2T {\sc apec} model fitted these spectra better than {\sc pshock} model
with a smaller value of $\rm {\chi_{\nu}^{2}}$ (see Figure~\ref{fig:mass_spt} and Table~\ref{tab:spec_massive_bestfit}).
The cooler and hotter X-ray temperatures deduced from X-ray spectra are found to be
respectively $\rm{0.62^{+0.02}_{-0.02}}$ keV and $\rm{>1.89}$ keV for DH Cep and
$\rm{0.32^{+0.07}_{-0.17}}$ keV and $\rm{0.74^{+0.13}_{-0.12}}$ keV for HD 97434, respectively.
The best fit values of $\rm {N_H}$ for  DH Cep and HD 97434
are in agreement with the values of $\rm {N_H}$ derived from optical data for
NGC 7380 ($\rm{0.32\times10^{22}~cm^{-2}}$ ; Massey, Johnson \& De Gioio-Eastwood 1995),
and Trumpler 18 ($\rm{0.14\times10^{22}~cm^{-2}}$ ; V$\rm{\acute{a}}$zquez \&
Feinstein 1990), respectively.
The $\rm{N_{H}}$ from optical data was estimated using the relation,
$\rm {N_H}$ $\rm{= 5.0\times}10^{21}\times{E(B-V)~cm^{-2}}$ (Vuong et al. 2003), where
$\rm{E(B-V)=A_V/3.1}$, assuming a normal interstellar reddening law towards the direction of the cluster. The extra absorbing hydrogen column density in addition to the interstellar contribution was not needed in the spectral fits implying  that  the absorption by the stellar wind is negligible (see also Zhekov \& Palla 2007; Sana et al. 2006).

\section{Discussion}\label{sec:discuss}

\subsection{X-ray Variability}
 The wind-shock model predicts a significant short-term variability in the X-ray flux as the shocks fade and grow on short time scales. However, it has not been observed so far.  We have also explored the possibility of short-term X-ray variability of DH Cep and HD 97434 over the observation time span ($<$40 ks).  The timing analysis reveals a lack of short term variability in the X-ray intensity for both the massive stars.
 The lack of short term variability could be due to the wind fragmentation, so that individual X-ray fluctuations are smoothed out over the whole emitting volume, leading to a rather constant output ( Feldmeier, Puls \& Pauldrach 1997).  Several other massive  stars, e.g., HD 159176 (De Becker et al. 2004) and HD 47129 (Linder et al. 2006) have also shown lack of variability in their light curves.

 The longer-term variability can not be probed using present data.  The long-term variability (few days) from O-type stars was reported by Snow, Cash \& Grady (1981) for the first time, but their results were  based on the low sensitivity detectors available at that time. No significant variability was detected by the ROSAT All-Sky Survey in the X-ray emission of 57 OB stars over a time-scale of $\approx$2 d (Bergh$\rm{\ddot{o}}$fer \& Schmitt 1995).

\subsection{X-ray temperatures of plasma}

Our analysis of XMM-Newton data reveals reveal two-component X-ray temperature structure for massive stars.  The temperatures corresponding to the cool and the hot components of HD 97434, and the cool component of DH Cep are found to be less than 1 keV. However, the temperature corresponding to the hot component of DH Cep was found to be more than 1.89 keV.  The derived values of temperatures are similar to  the O-type stars in the NGC~6231 cluster (Sana et al. 2006) and in the Carina OB1 association (Antokhin et al. 2008). Generally for O-type stars, the best fits to good quality data can usually be achieved by the sum of two thermal components at about 0.3 and 0.7--1\,keV (G${\rm{\ddot{u}}}$del \& Naz${\rm{\grave{e}}}$ 2009). The observed soft X-ray emission could originate from radiation-driven instabilities in stellar winds (Lucy 1982).  The wind-shock model predicts the intrinsic instability of the line driving mechanism.  Indeed, the velocity in an unstable wind is not the same everywhere and the fast-moving parcels of material will overcome the slow moving material, generating shocks between them.  This model  estimates the shock velocities (v$\rm{_{shock}}$)
by the relation  (Lucy 1982, Luo, McCray \& Maclow 1990)

\begin{equation}
\rm{v_{shock} =\sqrt{\frac{kT_{sh}}{1.95\mu}}}
\end{equation}

Adopting the mean particle weight
$\rm{\mu\approx}$0.62 for O-type stars (Cassinelli et al. 2008)
we derive the "average" value of shock velocities correspond to the cool components (kT$_1$)
$\rm{ 716^{+12}_{-12}  {~km~s^{-1} }}$ and $\rm{ 514^{+54}_{-162}  {~km~s^{-1} }}$
for the stars DH Cep and HD 97434, respectively.
However, the shock velocities correspond to the hot components (kT$_2$) are found to be
   $\rm{ >1563 {~km~s^{-1} }}$ and $\rm{ 782^{+66}_{-66}  {~km~s^{-1} }}$ for DH Cep and HD 97434, respectively.
These values are about a factor of 2 larger than the predicted by radiative shock  model
 of Lucy (1982).
 However, the evolved version of the standard model
by Owocki, Castor \& Rybicki (1988) predicts X-ray emission upto 1 keV from wind shocks model.
The temperature corresponding to the hot component of DH Cep is found to be more than 1.89 keV.
It may arise from wind collision zone of the binary system (see Bhatt et al. 2010 and references therein),
 but we are not able to draw any firm conclusion  on the basis of it alone.
Therefore, it appears that the cool as well as hot temperature components from DH Cep and HD 97434  could be
generated by instabilities in radiation-driven wind shocks.

\subsection{X-ray Luminosity}
For massive stars DH Cep and HD 97434, the ratios of X-ray to bolometric luminosities, $\rm{log(L_X /L_{bol}}$ ), are found to be -6.7 and -7.3 (see Table~\ref{tab:spec_massive_bestfit} ) in the energy band 0.3-7.5 keV, respectively, which are broadly consistent with the relation derived for similar kind of O-type stars (Sana et al. 2006).

X-ray luminosities derived from XMM-Newton data are found to be an order of magnitude lower than the $\rm{L_X}$
derived from Einstein observations for both the massive stars (see Table~\ref{tab:spec_massive_bestfit}
and \S\ref{sec:intro}).
These discrepancies can not be explained by difference in sensitivity ranges of the various instruments.
A similar results have been found by De Becker et al. (2004) for the star HD 159176 after the comparison of XMM-Newton, Einstein and ROSAT data.
We found the X-ray luminosity  $\rm{4.82\times10^{32}}$ and $\rm{1.95\times10^{31}}$ in energy range 0.3--2.0 keV for
the DH Cep and HD 97343, respectively, using XMM-Newton data.
The choice of the energy range is nearly a similar to the energy range
in Einstein observations, i.e., 0.2--3.5 keV. We further converted the Einstein IPC count rates into flux using the WebPIMMS\footnote{http://heasarc.gsfc.nasa.gov/Tools/w3pimms.html}, where we assumed a thermal plasma model with temperature of 0.5 keV, and the hydrogen column density of $3.9\times10^{21}$ for DH Cep and $1.8\times10^{21}$ for HD 97434 (see Table 2). The Einstein IPC count rates are 0.0166 and $<0.0103$ counts s$^{-1}$ for the stars DH Cep and HD 97343, respectively (Chlebowski et al. 1989).  The X-ray luminosity thus calulated was $\rm{2.3\times10^{32}}$ and $<\rm{1.2\times10^{31}}$ for the stars DH Cep and HD 97434, respectively. These value are consistent with that obtained from XMM-Newton obsertions. 
Therefore, it appears that the model used by Chlebowski et al. (1989)  to convert the  Einstein count rates into luminosities may be responsible for these discrepancies.

\section{Summary and Conclusions}
 Using the archival XMM-Newton observations, we present, for the first time,
the temporal and spectral analysis of X-ray emission from  O-type stars
DH Cep and HD 97434. The main results are as follows:
\begin{enumerate}
\item There is no firm evidence for short term variability in the X-ray intensity during the observations time span ($<$ 40 ks) for both the stars.

\item The values of $\rm{N_H}$ derived from the best fit X-ray spectra
are consistent with the optical estimates of reddening.

\item X-ray spectra of these stars are fitted consistently with two-temperature plasma models.
The X-ray emitting plasma is found to be generated at a temperature of lesser than 1 keV.
It could
originate from small shocks in the radiation-driven outflows. A hotter component is indicated in DH Cep.

\item The best-fit values of abundances are found to be 0.2 times of the solar abundances.
\end{enumerate}

\newpage

\begin{table}
\caption{Journal of XMM-Newton Observations of the objects.}             
\label{tab:xray_observations}      
\begin{tabular}{ccc}       
\hline
Object Name                         &  DH Cep                 & HD 97434              \\
\hline
Observation  ID                     &  0205650101             & 0051550101             \\
Exposure Time (s)                   &  31413                  & 40822                  \\
Start time UT (hh:mm:ss)            &  19 Dec  2003 02:02:12  &  06 Feb  2002 01:13:19 \\
Usable time(ks) (MOS1)$^{\dagger}$  &  25.23                  &                        \\
Usable time(ks) (MOS2)              &  25.46                  & 39.20                  \\
Usable time(ks) (PN)                &  19.52                  & 35.85                  \\  
EPIC filter                         &  Thick                  & Medium                 \\
Offset from Target (arcmin)         &  0.000                  & 3.68                    \\    
\hline
\end{tabular}
\newline
\noindent
$\dagger$ : The observations have not done in Prime Full window mode for MOS1 detector.
\end{table}

\begin{table*}
\caption {
The best-fit spectral parameters of the stars DH Cep and HD 97434.}
\label{tab:spec_massive_bestfit}      
\begin{tabular}{lllll}       
\hline
Parameters                                 & \multicolumn{2}{c} {DH Cep}                                                      &  \multicolumn{2}{c} {HD 97434}                      \\
                                           & \sc{pshock}                               &  2T \sc{apec}                               &  \sc{pshock}                                 & 2T \sc{apec}            \\
\hline
$\rm{N_H  (10^{22}~cm^{-2})}$              & $\rm{0.42^{+0.04}_{-0.04}}$           & $\rm{0.39^{+0.03}_{-0.03}}$            &  $\rm{0.17^{+0.04}_{-0.04}}$             & $\rm{0.18^{+0.05}_{-0.04}}$      \\
$\rm{kT_1  (keV)}$                         & $\rm{0.64^{+0.02}_{-0.02}}$           & $\rm{0.62^{+0.02}_{-0.02}}$            &  $\rm{0.64^{+0.04}_{-0.04}}$             & $\rm{0.32^{+0.07}_{-0.17}}$     \\
$\rm{kT_2  (keV)}$                         &                                       & $\rm{>1.89}$                           &                                          & $\rm{0.74^{+0.13}_{-0.12}}$      \\
$\rm{Z/Z_\odot}$                                   & $\rm{0.19^{+0.02}_{-0.02}}$           & $\rm{0.22^{+0.06}_{-04}}$              &  $\rm{0.19^{+0.07}_{-0.05}}$             & $\rm{0.21^{+0.11}_{-0.06}}$      \\
$\rm{\tau (10^{11}~s~cm^{-3})}$            & $\rm{15.14^{+59.94}_{-9.26}}$         &                                        &  $\rm{9.61^{+11.39}_{-4.23}}$            &                                  \\
$\rm{EM_1  (10^{54}~cm^{-3})}$             & $\rm{240.26^{+28.85}_{-33.79}}$       & $\rm{215.49^{+30.15}_{-43.86}}$        &  $\rm{5.09^{+1.62}_{-1.21}}$             & $\rm{4.09^{+3.18}_{-1.92}}$      \\
$\rm{EM_2  (10^{54}~cm^{-3})}$             &                                       & $\rm{4.95^{+9.28}_{-2.34}}$            &                                          & $\rm{3.43^{+2.02}_{-1.28}}$     \\
$\rm{\chi^{2}_{\nu}/d.o.f.}$               & 1.32(181)                             & 1.18(180)                              &  1.36(75)                                & 1.31(74)                        \\
$\rm{F_X (10^{-12}~erg~cm^{-2}~s^{-1})}$   & 0.33                                  & 0.34                                   &  0.08                                    &  0.08                           \\
log $\rm{L_X (erg~s^{-1})}$                & 32.73                                 & 32.75                                  &  31.31                                   &  31.32                          \\
$\rm{log(L_X/L_{bol})[0.3-7.5]}$           & -6.73                                 &  -6.71                                 &  -7.29                                   &  -7.29                          \\
$\rm{log(L_X/L_{bol})[0.3-2.0]}$           & -6.77                                 &  -6.77                                 &  -7.31                                   &  -7.31                          \\
$\rm{log(L_X/L_{bol})[2.0-7.5]}$           & -7.76                                 &  -7.55                                 &  -8.56                                   &  -8.48                          \\
\hline
\end{tabular}
\end{table*}

\clearpage

\end{document}